\documentclass[letterpaper]{physor2020}
\usepackage{tabls}
\usepackage{cites}
\usepackage{epsf}
\usepackage{appendix}
\usepackage{ragged2e}
\usepackage[top=1in, bottom=1.in, left=1.in, right=1.in]{geometry}
\usepackage{enumitem}
\setlist[itemize]{leftmargin=*}
\usepackage{caption}
\usepackage{amssymb}
\usepackage{amsthm,amssymb, amsfonts, amsmath}
\usepackage{units}
\usepackage{graphicx}
\usepackage{listings}
\usepackage{multirow}
\usepackage{booktabs}
\usepackage{float}
\floatstyle{plaintop}
\restylefloat{table}
\usepackage{booktabs}
\usepackage{enumerate}
\graphicspath{{figures/}}
\usepackage{caption}
\usepackage{subcaption}

\title{Fast Rossi-alpha Measurements of Plutonium using Organic Scintillators}
\author{%
  \textbf{M.Y. Hua$^{1,2}$, C.A. Bravo$^1$, A.T. MacDonald$^1$, J.D. Hutchinson$^2$, G.E. McKenzie$^2$,} \\ 
  \textbf{T.J. Grove$^2$, J.M. Goda$^2$, A.T. McSpaden$^2$, S.D. Clarke$^1$, and S.A. Pozzi$^1$} \\
  \\
  $^1$Department of Nuclear Engineering and Radiological Sciences, University of Michigan,  \\
  2355 Bonisteel Blvd, Ann Arbor, MI 48109, USA \\ 
\\
  $^2$NEN-2: Advanced Nuclear Technology Group, Los Alamos National Laboratory,   \\ 
    PO Box 1663, MS B228, Los Alamos, NM 87545, USA \\ 
    \\
  \textit{\url{mikwa@umich.edu}}
}
\newcommand{\authorHead}      
           {Hua et al.}  
\newcommand{\shortTitle}      
           {Fast Rossi-alpha Measurements of Plutonium using Organic Scintillators}  
\begin{document}
\maketitle
\justify 

\begin{abstract}
  In this work, Rossi-alpha measurements were simultaneously performed with a $^3$He-based detection system and an organic scintillator-based detection system.   The assembly is 15 kg of plutonium (93 wt\% $^{239}$Pu) reflected by copper and moderated by lead.  The goal of Rossi-alpha measurements is to estimate the prompt neutron decay constant, alpha.  Simulations estimate $k_\text{eff}=0.624$ and $\alpha=52.3\pm 2.5$ ns for the measured assembly.  The organic scintillator system estimated $\alpha = 47.4 \pm 2.0$ ns, having a 9.37\% error (though the 1.09 standard deviation confidence intervals overlapped).  The $^3$He system estimated $\alpha = 37$ $\mu$s.  The known slowing down time of the $^3$He system is 35-40 $\mu$s, which means the slowing down time dominates and obscures the prompt neutron decay constant.  Subsequently, the organic scintillator system should be used for assemblies with alpha much less than 35 $\mu$s.
\end{abstract}

\section{INTRODUCTION AND MOTIVATION} 
Reactivity estimates are an important facet of nuclear criticality safety.  Currently, reactivity cannot be directly measured and is instead inferred from the prompt neutron decay constant, $\alpha$.  The value of $\alpha$ is estimated using Rossi-alpha measurements that are predicated on measuring the time difference between neutron detections~\cite{uhrig}.  Rossi-alpha measurements are traditionally conducted with $^3$He detectors, which use polyethylene to moderate neutrons to improve the detection efficiency.  Because neutrons take time to moderate in the polyethylene, timing properties change (the decay of prompt neutrons and neutron slowing down time are convolved) and information can be lost.  Organic scintillation detectors can detect neutrons directly and are fast compared to $^3$He systems.  In this work, a fast plutonium assembly is simultaneously measured by $^3$He detectors and organic scintillators, and the detection systems are compared.

\section{THE ROSSI-ALPHA METHOD}
\begin{figure}[H]
	\centering
	\includegraphics[width=.45\linewidth]{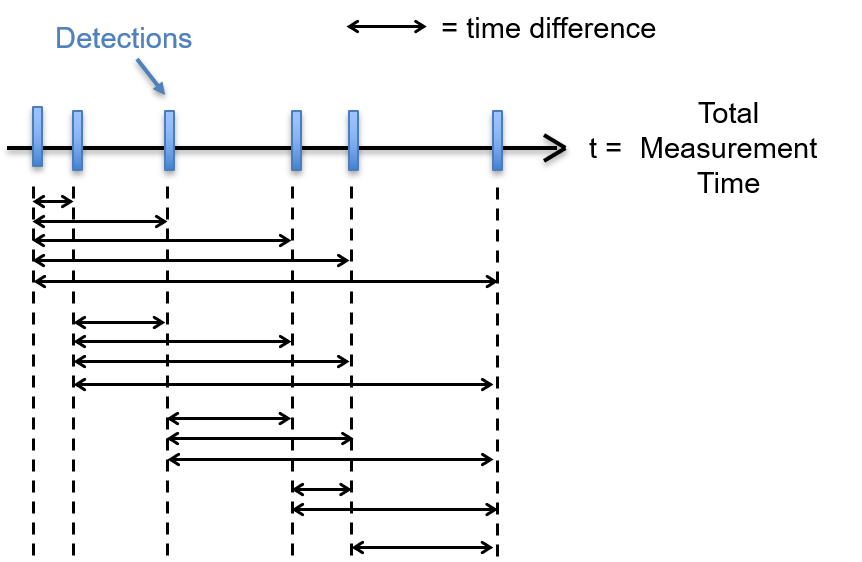}
	\caption{Time difference calculation for Rossi-alpha measurements.}
	\label{fig:RA} 
\end{figure}

\begin{figure}[H]
	\centering
	\includegraphics[width=.87\linewidth]{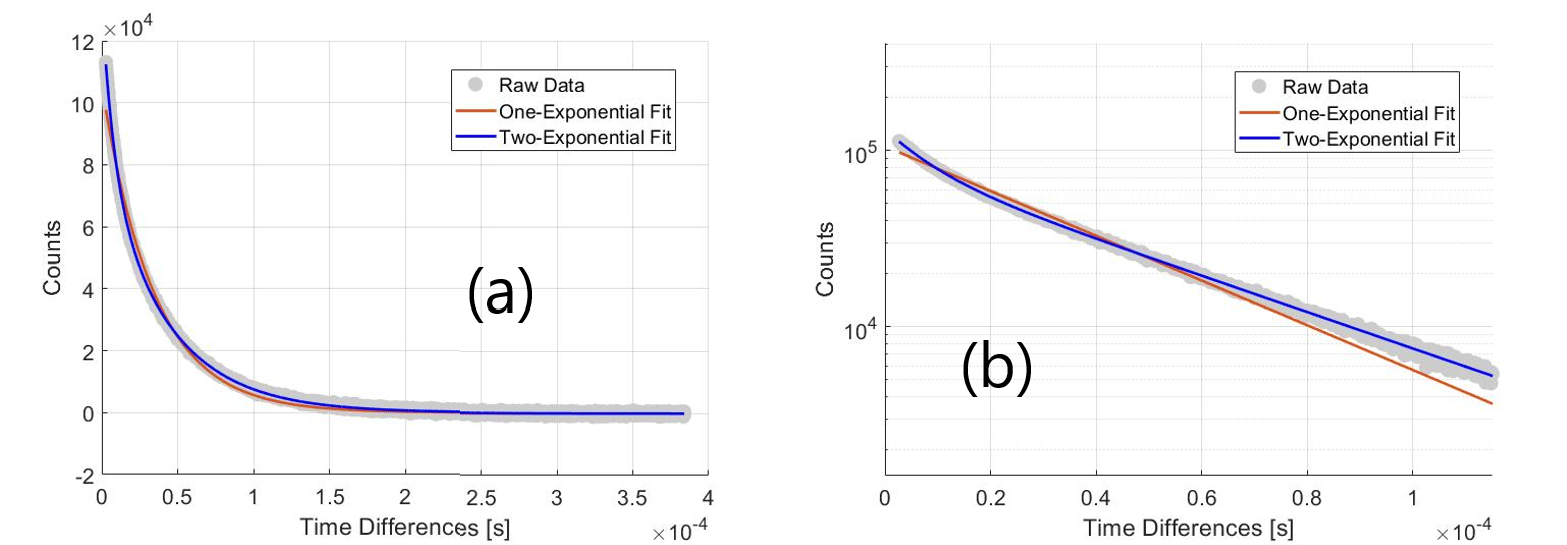}
	\caption{Rossi-alpha plot with one- and two-exp fits on linear (a) and semilog (b) scales.}
	\label{fig:sample_RA}
\end{figure}

In a Rossi-alpha measurement, neutron detection times are recorded, the time differences between detections are calculated (see Fig.~\ref{fig:RA}), and a Rossi-alpha histogram of the time differences is constructed~\cite{uhrig,Feynman44_1,hansen}.  The histogram is fit and $\alpha$ is obtained from the fit parameters.  Traditionally, the histogram is fit with a single-exponential-plus-constant model \cite{ornbro}.  Recent work has developed a double-exponential-plus-constant-model~\cite{mikwa_2exp} that is more suitable than the single exponential model for measurements of reflected assemblies.  The two-exponential model estimates a second parameter: $\ell_{ctd}$, which describes the mean time a neutron spends in the reflector prior to detection.  Sample Rossi-alpha histograms from a $^3$He detector measuring plutonium are shown in Fig.~\ref{fig:sample_RA}.
\section{ASSEMBLY SPECIFICATIONS AND EXPERIMENTAL SETUP}
\subsection{Assembly Specifications}
\begin{figure}[H]
	\centering
	\includegraphics[width=.45\linewidth]{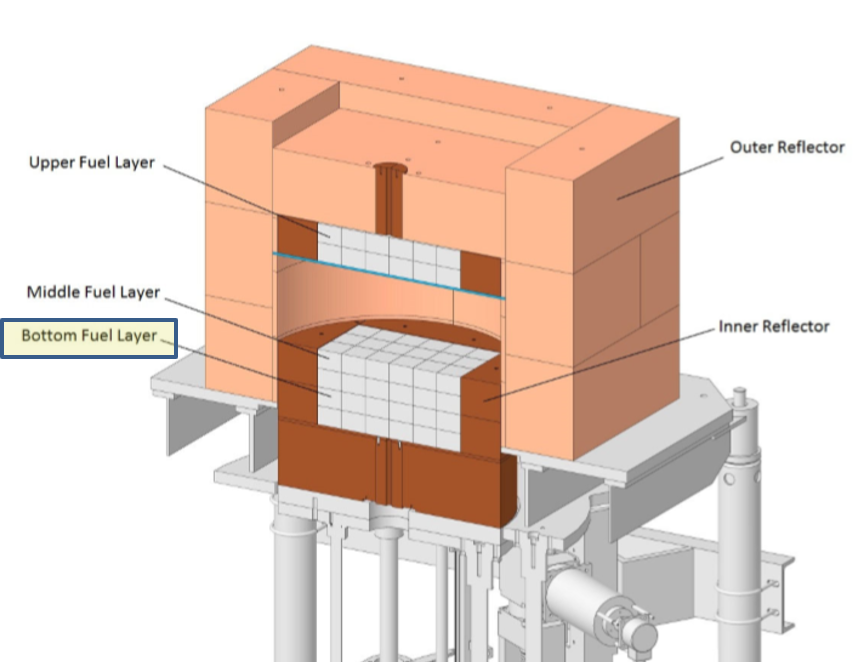}
	\caption{3D rendering of the Comet critical assembly~\cite{joetta_PHYSOR}.}
	\label{fig:3D}
\end{figure}

\begin{figure}[H]
	\centering
	\begin{minipage}{.5\textwidth}
		\centering
		\includegraphics[width=.7\linewidth]{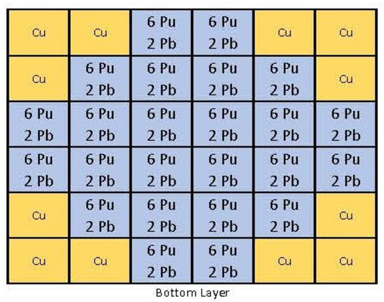}
		\caption{Bottom layer box layout~\cite{joetta_PHYSOR}.}
		\label{fig:bottom_layer}
	\end{minipage}%
	\begin{minipage}{.5\textwidth}
		\centering
		\includegraphics[width=.6\linewidth]{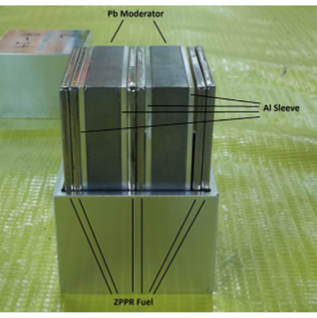}
		\caption{Photo of a plutonium box \cite{joetta_PHYSOR}.}
		\label{fig:box}
	\end{minipage}
\end{figure}

In this work, the bottom layer of the Comet critical assembly -- lead-moderated, copper-reflected plutonium (93 wt\% $^{239}$Pu) -- was measured.  A 3D rendering of the assembly is shown in Fig.~\ref{fig:3D}; the layout of the bottom layer of copper or plutonium boxes is shown in Fig.~\ref{fig:bottom_layer}, and a sample plutonium box is shown in Fig.~\ref{fig:box} ~\cite{joetta_PHYSOR}.  The total mass of plutonium was approximately 15 kg.  
\subsection{Simulation of the Assembly}
\begin{figure}[H]
	\centering
	\includegraphics[width=.45\linewidth]{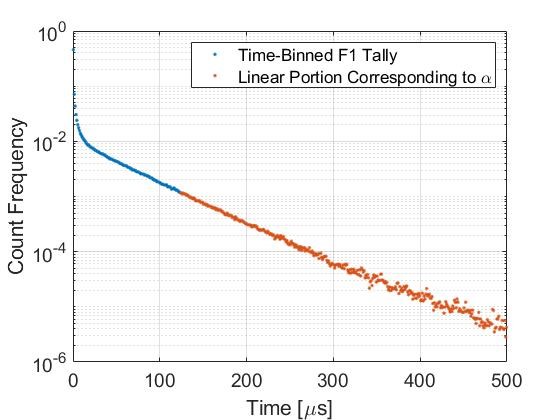}
	\caption{Sample plot of the time-binned surface tally (F1) used to estimate the Rossi-alpha.}
	\label{fig:TiBTaF}
\end{figure}
To estimate the prompt neutron decay constant, $\alpha$, the measurement was simulated using MCNP6\textregistered.  The KCODE option estimated $k_\text{eff} \approx 0.624$.  To determine $\alpha$, surface (F1) and point-detector (F5) tallies were time-binned, and the tails (linear on a semilog plot) were fit.  A sample time-bin tail-fit plot is shown in Fig.~\ref{fig:TiBTaF} and $\alpha = 52.3\pm2.5$ ns.  The uncertainty comes from the fit uncertainty.  
\subsection{Experimental Setup and Detection System Details}
\begin{figure}[H]
	\centering
	\begin{minipage}{.5\textwidth}
		\centering
		\includegraphics[width=.9\linewidth]{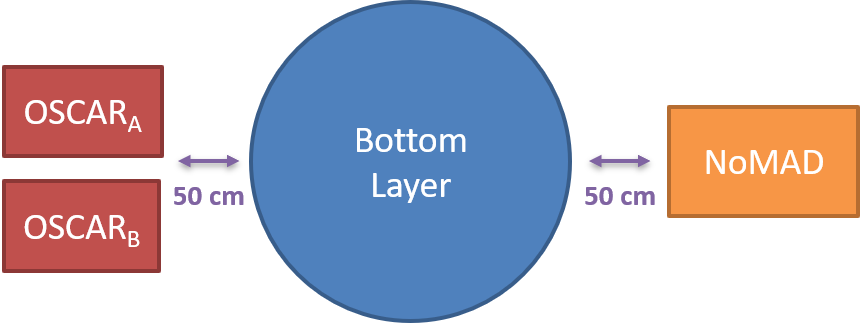}
		\caption{Schematic of detector placement.}
		\label{fig:schematics}
	\end{minipage}%
	\begin{minipage}{.5\textwidth}
		\centering
		\includegraphics[width=.9\linewidth]{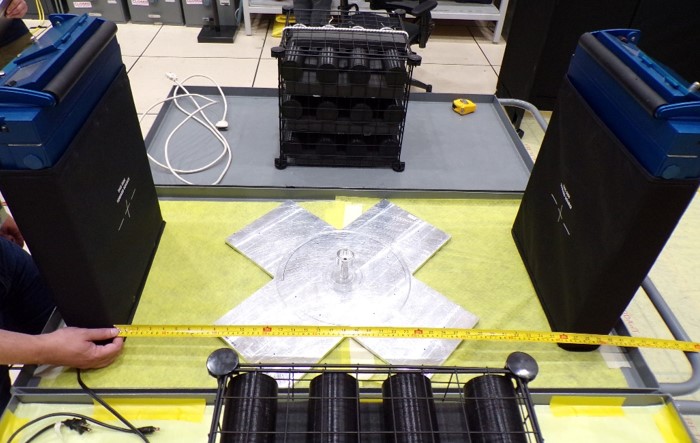}
		\caption{Photo of detection systems.}
		\label{fig:photo}
	\end{minipage}
\end{figure}

In the measurement of the assembly, two organic scintillator arrays (OSCARs) and one Neutron Multiplicity $^3$He Array Detector (NoMAD) were used. An OSCAR comprises 12 5.08 cm $\times$ 5.08 cm diameter \textit{trans}-stilbene organic scintillators coupled to photomultiplier tubes~\cite{stilbene,stilbene2}.  The NoMAD is similar to the MC-15 detection system \cite{mc15_manual}, comprising 15 $^3$He detectors embedded in a polyethylene matrix.  The systems were placed 50 cm from the edge of the assembly; a schematic is shown in Fig.~\ref{fig:schematics} and a photo of the systems side-by-side is shown in Fig.~\ref{fig:photo}.  For this work, only 21 of the 24 OSCAR detectors were operational.  Based on neutron detection rates, the NoMAD (in the given configuration) is 3.34 times more efficient than the OSCARs.      
\section{DATA ANALYSIS}
\subsection{Data Analysis for the $^3$He-based NoMAD System}\label{sec:DA_NoMAD}
\begin{figure}[H]
	\centering
	\includegraphics[width=.5\linewidth]{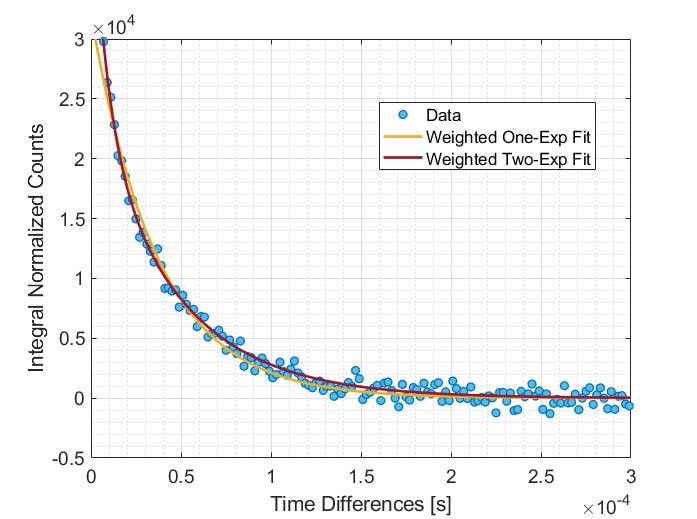}
	\caption{Rossi-alpha histogram with fits from the NoMAD system.}
	\label{fig:NoMAD_RA}
\end{figure}

The output from measurement and preliminary data analysis is a list of detection times.  The Rossi-alpha histogram is created using type-I binning (illustrated in Fig.~\ref{fig:RA})~\cite{hansen}.  In theory, Rossi-alpha histograms peak at a time difference of 0 s; however, measurement considerations such as dead time and time of flight cause the peak to occur later.  Suppose the max occurs in bin $b$.  To mitigate the measurement considerations at short time differences, the first $2b$ bins are discarded.  For some comparison purposes in this work, the histograms are integral normalized and constant-subtracted by taking the mean of the last points in the tail.  A sample, resultant Rossi-alpha histogram for the NoMAD is shown in Fig.~\ref{fig:NoMAD_RA}.
\subsection{Data Analysis for the Organic Scintillator-based OSCAR System}
\begin{figure}[H]
	\centering
	\begin{minipage}{.5\textwidth}
		\centering
		\includegraphics[width=.9\linewidth]{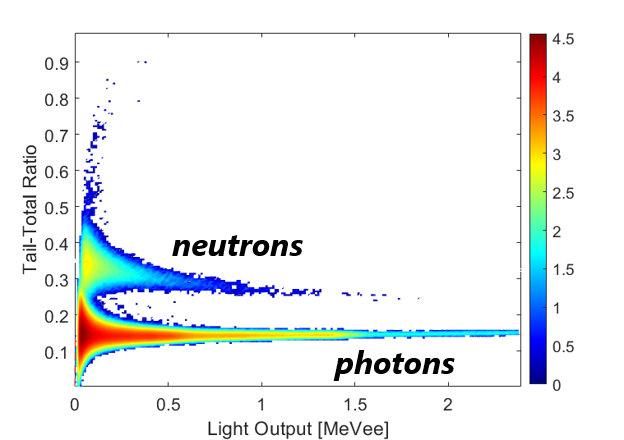}
		\caption{Sample PSD plot.}
		\label{fig:PSD}
	\end{minipage}%
	\begin{minipage}{.5\textwidth}
		\centering
		\includegraphics[width=.9\linewidth]{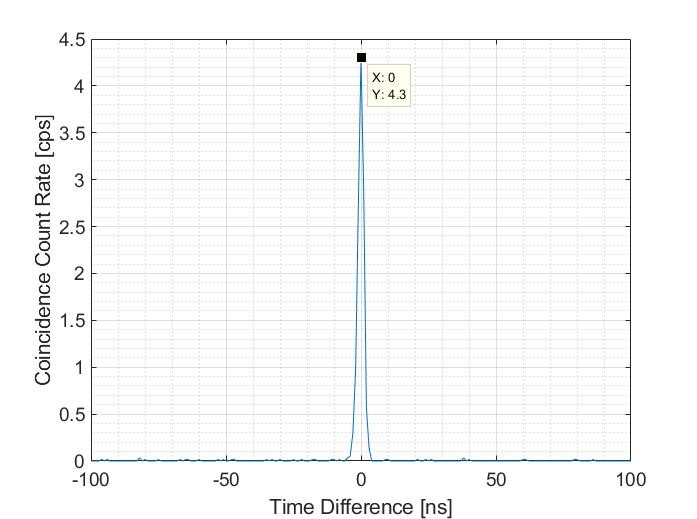}
		\caption{Sample photon-photon coincidence plot.}
		\label{fig:PP_offset}
	\end{minipage}
\end{figure}

\begin{figure}[H]
	\centering
	\includegraphics[width=.5\linewidth]{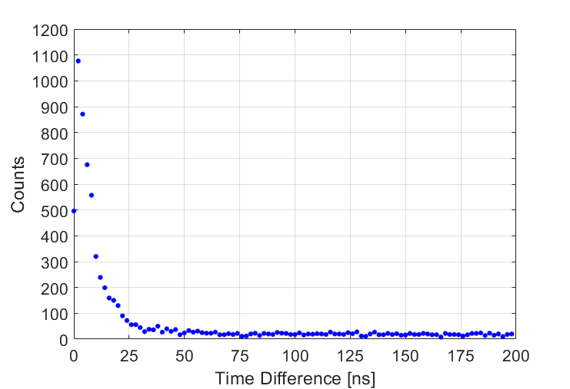}
	\caption{Rossi-alpha histogram from the OSCAR system.}
	\label{fig:OSCAR_RA}
\end{figure}

The output from measurement and preliminary data analysis is a list of detection times, total pulse integrals, and tail integrals.  Because organic scintillators are sensitive to both neutrons and photons, pulse shape discrimination (PSD) is used to discriminate the pulses.  The PSD is done for each detector and is both time and energy dependent; a sample PSD plot is shown in Fig.~\ref{fig:PSD}.  The PSD analysis results in three sets of data: neutron pulses, photon pulses, and pulses to discard (due to, for example, pulse pileup).  Currently, gamma-ray Rossi-alpha is not considered; however, the photon pulses are still needed to correct for timing offsets. The OSCAR system is sensitive to time of flight and offsets due to electronics.  To correct for offsets, the photon-photon coincidence peak (shown in Fig.~\ref{fig:PP_offset}, present from prompt fission photons) is created for all detectors relative to one detector.  If the peak is not centered about zero, all times in the neutron and photon pulse lists are subsequently shifted by a constant. Once the list of neutron detection times is corrected, the Rossi-alpha analysis is the same as that for the NoMAD system (see Section~\ref{sec:DA_NoMAD}); the resultant Rossi-alpha plot is shown in Fig.~\ref{fig:OSCAR_RA}.  
\section{RESULTS AND DISCUSSION}
\begin{figure}[H]
	\centering
	\includegraphics[width=\linewidth]{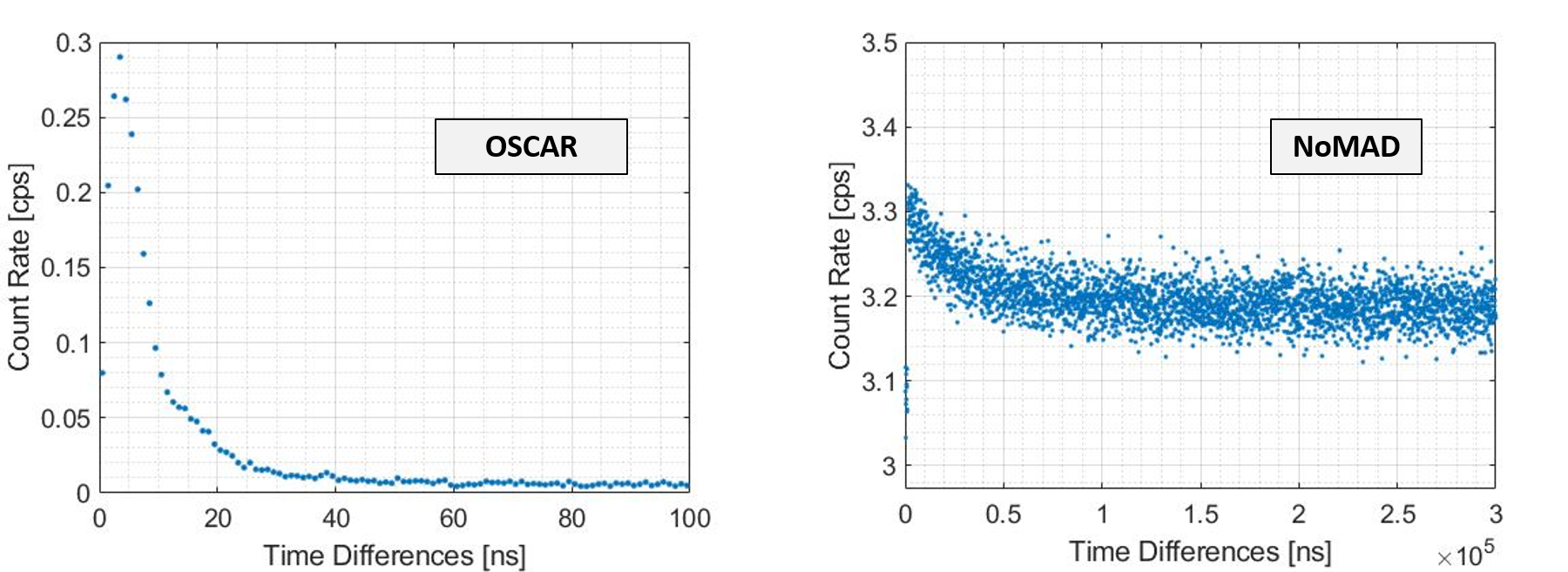}
	\caption{Unnormalized, non-constant-subtracted Rossi-alpha histograms.}
	\label{fig:accidentals}
\end{figure}

Unnormalized, non-constant-subtracted Rossi-alpha histograms generated from two hours of data for each detection system are shown in Fig.~\ref{fig:accidentals}.  The OSCAR has fewer accidentals than that of the NoMAD: the constant value of the tail for the NoMAD is 95\% of the maximum value, whereas the constant value of the tail for the OSCAR is only 0.7\% of the maximum value.  In some cases, the high proportion of the accidentals in the case of the NoMAD may obscure the second exponential.  Obscuring the second exponential would reduce the fit model to a single exponential fit; however, since the parameters of interest are a linear combination of the two exponentials, $\alpha$ and $\ell_{ctd}$ cannot be determined.

\begin{figure}[H]
	\centering
	\includegraphics[width=\linewidth]{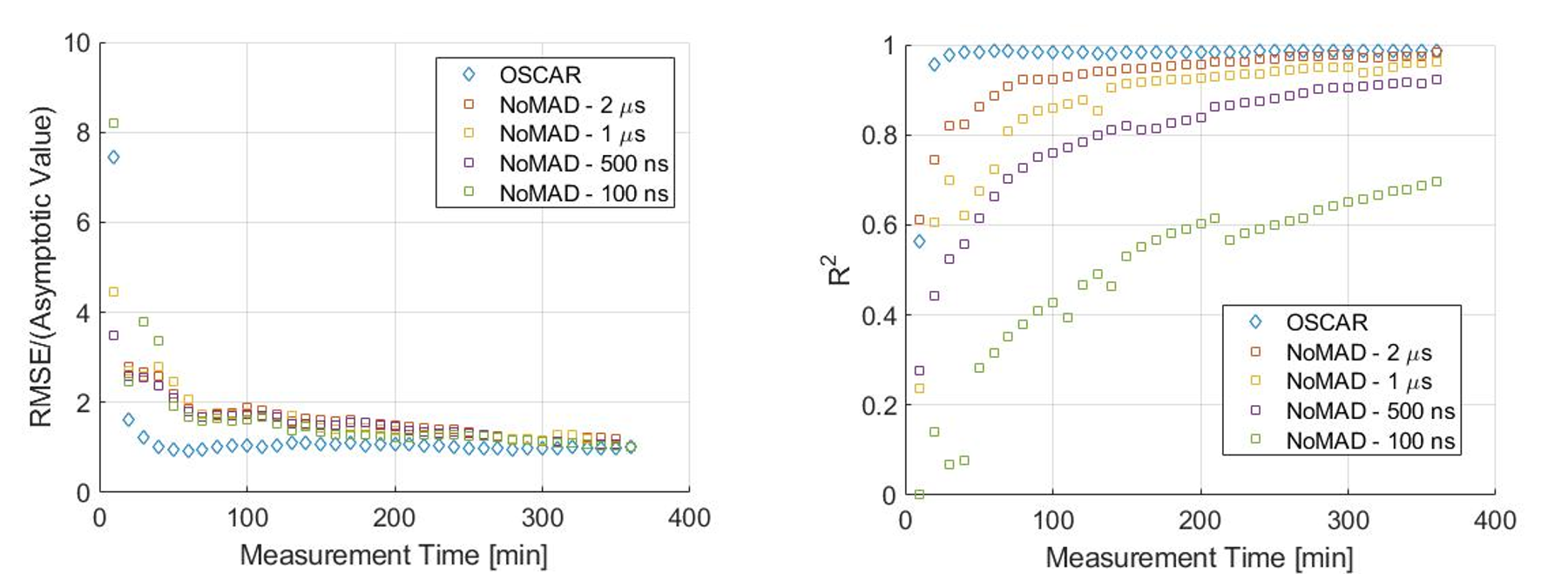}
	\caption{Fit metrics as a function of measurement time for the NoMAD at different bin widths and the OSCAR.}
	\label{fig:fit_metrics}
\end{figure}

Fit metrics plotted as a function of measurement time (and bin width for the NoMAD) are shown in Fig.~\ref{fig:fit_metrics}.  The root mean square error (RMSE) is normalized by the asymptotic values of the respective data series such that the y-axis is a measure of convergence.  It takes the OSCAR less than 30 minutes to be within 50\% of its asymptotic value, while it takes the NoMAD approximates 120 minutes (note that RMSE is fairly independent of the bin width, as expected).  It takes the OSCAR less than 20 minutes to achieve an $R^2$ value greater than 0.90, whereas the the NoMAD with 2 $\mu$s bins requires approximately 70 minutes.  The NoMAD's $R^2$ convergence could be improved by increasing the bin widths; however, 2 $\mu$s bin widths are already large compared to the time-decay constant (52.3 $\pm$ 2.5 ns) the NoMAD is trying to observe.  Reducing the bin widths to increase sensitivity to the physical phenomenon the system is trying to measure results in increases in the time is takes the NoMAD to achieve $R^2 > 0.90$; bin widths of 1 $\mu$s require 140 minutes and bin widths of 500 ns require 280 minutes (the relationship is approximately linear).   

From simulation, the ``true" value of $\alpha$ for the assembly is taken to be 52.3 $\pm$ 2.5 ns.  Fitting the OSCAR data with a two exponential, $\alpha$ is estimated to be 47.4 $\pm$ 2.0 ns.  The error is 9.37\% and, qualitatively, the values are similar since the $1.09\sigma$-confidence intervals overlap.  The NoMAD estimate of $\alpha$ is $\approx 37$  $\mu$s.  The NoMAD has a known slowing down time of 35-40 $\mu$s and, because $\alpha\ll 35$ $\mu$s, the NoMAD is likely only sensitive to the neutron moderation time. 

\section{CONCLUSIONS AND FUTURE WORK}
In this work, the organic scintillator array (OSCAR), comprising 21 total operational \textit{trans}-stilbene detectors, and the Neutron Multiplicity $^3$He Array Detector (NoMAD), comprising 15 $^3$He tubes embedded in a polyethylene matrix, simultaneously measured 15 kg of plutonium (93 wt\% $^{239}$Pu) moderated by lead and reflected by copper with $k_\text{eff} = 0.624$ and $\alpha = $ 52.3$\pm$2.5 ns.  It was found that the OSCAR converged on its estimate of $\alpha$ faster than the NoMAD, which translates to reduced procedural and operational costs in practical implementation.  The convergence needs to be investigated further for assemblies where $\alpha$ is much larger ($\alpha\propto 10-100s$ of $\mu$s). Because neutrons are moderated in the polyethylene matrix of the NoMAD (and moderation is not inherent to the OSCAR), the OSCAR is an inherently faster detection system.  The entire Rossi-alpha histogram (reset time) is less than 100 ns for the OSCAR (1 ns bins), whereas 100 ns is the clock tick length for the NoMAD.  Therefore, for fast assemblies ($\alpha \propto 1-100s$ of ns), it is more suitable to use the OSCAR that estimated the true $\alpha$ within 1.09 standard deviations and an error of 9.37\% (on the order of uncertainty in nuclear data).  Larger accidental contributions are more likely to wash out time information; the NoMAD has a large accidental contribution and the OSCAR has a negligible accidental contribution. Future work involves determining when each system is more suitable to a given measurement.  Furthermore, gamma-ray and mixed-particle Rossi-alpha will be investigated with the organic scintillators.  Work will also be done with more measurements to validate organic scintillator-based Rossi-alpha measurements.

\section*{ACKNOWLEDGEMENTS}

This work is supported by the National Science Foundation Graduate Research Fellowship under Grand No. DGE-1256260, by the Consortium for Verification Technology under Department of Energy National Nuclear Security Administration award number DENA0002534, and by the DOE Nuclear Criticality Safety Program, funded and managed by the National Nuclear Security Administration for the Department of Energy.  Any opinion, findings, and conclusion or recommendations expressed in this material are those of the authors and do not necessarily reflect the views of the National Science Foundation or other funding organizations.


\setlength{\baselineskip}{12pt}
\bibliographystyle{physor}
\bibliography{bibliography}

\begin{thebibliography}{1}
\newcommand{\enquote}[1]{``#1''}
\providecommand{\url}[1]{\texttt{#1}}
\providecommand{\urlprefix}{URL }

\bibitem{uhrig}
R.~Uhrig and U.~A.~E. Commission.
\newblock \emph{Random noise techniques in nuclear reactor systems}.
\newblock Ronald Press (1970).

\bibitem{Feynman44_1}
R.~Feynman, F.~DeHoffmann, and R.~Serber.
\newblock \enquote{Statistical Fluctuations in the Water Boiler and the
  Dispersion of Neutrons Emitted per Fission.}
\newblock \emph{LA-101, Los Alamos National Laboratory} (1944).

\bibitem{hansen}
G.~E. Hansen.
\newblock \enquote{The Rossi alpha method.}
\newblock \emph{Los Alamos National Laboratory, Technical Report},
  (LA-UR-85-4176) (1985).

\bibitem{ornbro}
J.~Orndoff.
\newblock \enquote{Prompt Neutron Periods of Metal Critical Assemblies.}
\newblock \emph{Nuclear Science and Engineering}, \textbf{volume~2}(4), pp.
  450--460 (1957).

\bibitem{mikwa_2exp}
M.~Hua, J.~Hutchinson, G.~McKenzie, T.~Shin, S.~Clarke, and S.~Pozzi.
\newblock \enquote{Derivation of the Two-Exponential Probability Density
  Function for Rossi-alpha Measurements of Reflected Assemblies and Validation
  for the Special Case of Shielded Measurements.}
\newblock \emph{DOI: 101080/0029563920191654327} (2019).

\bibitem{joetta_PHYSOR}
J.~Goda, T.~Cutler, T.~Grove, J.~Hutchinson, G.~McKenzie, A.~McSpaden,
  M.~Nelson, and R.~Sanchez.
\newblock \enquote{Comparison of Methods for Determining Multiplication in
  Subcritical Configurations of a Plutonium System.}
\newblock \emph{Proceedings of the PHYSOR Conference} (2018).

\bibitem{stilbene}
N.~Zaitseva, A.~Glenn, L.~Carman, H.~P. Martinez, R.~Hatarik, H.~Klapper, and
  S.~Payne.
\newblock \enquote{Scintillation properties of solution-grown trans-stilbene
  single crystals.}
\newblock \emph{Nuclear Instruments and Methods in Physics Research Section A:
  Accelerators, Spectrometers, Detectors and Associated Equipment},
  \textbf{volume 789}, pp. 8 -- 15 (2015).

\bibitem{stilbene2}
M.~Bourne, S.~Clarke, N.~Adamowicz, S.~Pozzi, N.~Zaitseva, and L.~Carman.
\newblock \enquote{Neutron detection in a high-gamma field using solution-grown
  stilbene.}
\newblock \emph{Nuclear Instruments and Methods in Physics Research Section A:
  Accelerators, Spectrometers, Detectors and Associated Equipment},
  \textbf{volume 806}, pp. 348 -- 355 (2016).

\bibitem{mc15_manual}
C.~Moss, M.~Nelson, R.~Rothrock, and E.~Sorensen.
\newblock \enquote{MC-15 Users Manual.}
\newblock \emph{Los Alamos National Laboratory, Technical Report},
  (LA-UR-18-29563) (2018).

\end{thebibliography}

\end{document}